\documentclass[3p,authoryear]{elsarticle}
\usepackage{amsmath}
\usepackage{amssymb}
\usepackage{amsthm}
\usepackage{mathtools}
\usepackage{hyperref}
\usepackage{graphicx}
\usepackage{xparse}
\usepackage{tikz-feynman}
\usepackage{enumitem}

\DeclareDocumentCommand \fundiff{m}{\mathcal{D}#1}

\DeclareDocumentCommand \funint{> { \SplitList { , } } m }{\int\ProcessList{#1}{\fundiff}}

\newcommand{\fundelta}[2]{\frac{\delta#1}{\delta#2}}

\newcommand{\Tr}{\textnormal{Tr}}

\newcommand{\Lgr}{\mathcal{L}}

\numberwithin{equation}{section}

\journal{Nuclear Physics B}

\begin{document}

\unitlength = 1mm

\begin{frontmatter}

\title{A possibility for grand unification and non-Higgs mass generation in a Nambu-Jona-Lasinio-like theory of fermions interacting with current metric field}
\author{Sergii Kutnii}

\begin{abstract}
Grand unification possibilities in Nambu-Jona-Lasinio-like models are studied. 
To address the problem of vector boson masses and nonrenormalizability of the theory, 
algebraic formalism encompassing the effective action, Schwinger-Keldysh path integral, 
and Bogoliubov-Parasiuk-Hepp-Zimmerman renormalization is constructed. 
A new NJL-like model: the theory of current metric field interacting with fermions is proposed.
Bosonization in this model can produce massless vector bosons under certain
conditions which makes it a candidate grand unified theory. 
Both Higgs and non-Higgs effects can contribute to particle masses.
\end{abstract}


\end{frontmatter}




\section{Introduction}

As almost a decade of observations at Super-Kamiokande has set lower limit on proton lifetime to $\sim 10^{34}$ years,
most simple grand unification models based on spontaneous symmetry breaking in non-supersymmetric Yang-Mills theories have been ruled out \cite{SuperKamiokande2016}.
At the same time, supersymmetry has not been observed yet.

While there still remain non-supersymmetric gauge candidates for grand unification \cite{Ellis2002}, \cite{BabuKhan2015}, 
non-renormalizable alternaives are also worth exploring. 
One such model of Nambu-Jona-Lasinio (NJL) \cite{NJL} type was proposed by Terazawa, Chikashige, and Akama \cite{Terazawa1977}.
They demonstrated that bosonization in a theory with four-fermion interactions may produce all the observable particle types.
The biggest difficulty in such a program is that vector bosons emerge naturally massive in it. 
The authors suggested that masslessness of the photon resulted from accidental cancellations in the effective theory. Conceptually similar models were studied in \cite{Eguchi1974}, \cite{Kawasaki1981}.

The problem with the aforementioned approach is that if the theory is merely regularized which seems to be the default method for a non-renormalizable theory such as the NJL one,
cancellations will depend on the regularization. In particular, under the momentum cutoff at $\Lambda$, the bubble diagram
\begin{equation}
	\begin{tikzpicture}[baseline=(b)]
		\begin{feynman}
			\vertex(a);
			\vertex [right=0.5cm of a] (b);
			\vertex [right=1cm of b] (c);
			\vertex [right=0.5cm of c] (d);
			\diagram* {
				(a) --[photon, very thick] (b),
				(b) --[fermion, half right] (c),
				(c) --[fermion, half right] (b),
				(c) --[photon, very thick] (d)
			};
		\end{feynman}
	\end{tikzpicture}
\end{equation}
gives a contribution to vector boson mass with the leading order of $\Lambda$. 
In gauge theories, this is considered an artifact of cutoff regularization. 
Indeed, this contribution does not appear in the dimensional or Pauli-Villars regularizations.
Thus, the question "what if cancellation is merely a regularization artifact?" would plague models of the NJL type 
without a more robust approach to divergences. 

It is not hard to notice that effective action formalism can yield the same results as the usual Hartree-Fock-Bogoliubov treatment of NJL-like models while 
allowing to study higher loop corrections.
Thus, Jackiw's 1974 idea: to apply Bogoliubov's R-operation to the effective action \cite{Jackiw1974} suggests a way forward as R-operation formalism can in principle be applied to non-renormalizable theories at the cost of introducing infinitely many renormalization coinstants. The original R-operation scheme, however, was formulated in terms of operator quantum field theory and relied on certain assumptions about the vacuum. The problem is that vacuum is not known a priori in the effective action formalism. Instead, possible vacua arise as solutions to the equation
\begin{equation}
\fundelta{\Gamma}{\varphi} = 0
\end{equation}
where $\Gamma$ is the effective action and $\varphi$ is the mean field. It would be natural to assume that particles arise as small deviations of the field from the background but as Haag's results \cite{Haag1955} suggest, different backgrounds can give rise to totally different Hilbert spaces of particle states. Thus, an attempt to apply the original R-operation formalism to the effective action leads to a vicious circle where the vacuum cannot be determined without the effective action and the latter could not be properly renormalized without knowing the vacuum. Fortunately, modern algebraic quantum field theory \cite{Haag2012}, \cite{Duetsch2001}, \cite{DuetschFred2004}, \cite{Duetsch2005}, \cite{FredRejz2012}, \cite{HawkinsRejzner2019}, \cite{Brunetti2022} allows a purely functional formulation of the Bogoliubov-Parasiuk-Hepp-Zimmermann (BPHZ) renormalization, avoiding the complications arising in the operator approach.

Thus, the problem of integrating the effective action formalism into the framework of algebraic quantum field theory arises naturally.
This problem is addressed in the section \ref{formalism}.
In the section \ref{NJL-effaction}, the properties of renormalized effective action for an NJL-like model are investigated,
and in \ref{current_metric}, the current metric field is introduced and it is demonstrated how a generalized NJL model with 
current metric field can produce a reasonable grand unified theory.
Implications of the study are discussed in the conclusions and discussion section.

\section{Effective action in algebraic quantum field theory} \label{formalism}

Algebraic quantum field theory starts with the notion of an involutive complex algebra $\mathfrak{U}$ \cite{FredRejz2012} 
i.e. an algebra over the field of complex numbers equipped with a map $^\dag:\mathfrak{U} \rightarrow \mathfrak{U}$ such that
\begin{equation}
\begin{split}
&\forall A,B \in \mathfrak{U} : \left(A+B\right)^{\dag} = A^{\dag} + B^{\dag},(AB)^{\dag} = B^{\dag}A^{\dag}\\
&\forall \lambda \in \mathbb{C}, A \in \mathfrak{U} : \left(\lambda{A}\right)^{\dag} = \bar{\lambda}A^{\dag}\\
&\forall A \in \mathfrak{U} : (A^{\dag})^{\dag} = A
\end{split} \label{formalism:defalgebra}
\end{equation}

Let $\phi_a(x) : a \in \left\{1,2\right\}$ be a possibly complex field vaiable (the meaning of the index $a$ will be clarified below). While bosonic field is assumed for simplicity, the formalism can be extended to fermions along the lines of \cite{Brunetti2022}. Let $\mathfrak{X} = \left\{\chi_i\left[\phi\right]\right\}$ be some set of linearly independent infinitely differentiable functionals in $\phi$ (composite fields) such that $\forall (x,a) : \phi_a(x) \in \mathfrak{X}$ and the index $i$ is an abstract possibly continuous index.
The algebra $\mathfrak{U}$ is the polynomial field generated by elements of $\mathfrak{X}$. The involution $^{\dag}$ acts on the fields as follows:
\begin{equation}
\begin{split}
&\left[\phi_1(x)\right]^{\dag} = \bar{\phi}_2(x)\\
&\left[\phi_2(x)\right]^{\dag} = \bar{\phi}_1(x)
\end{split}
\end{equation}

Any element in $\mathfrak{U}$ can be written as
\begin{equation}
\begin{split}
&F\left[\phi\right] = \sum{}F_{i_1\ldots{i_n}}\chi_{i_1}\left[\phi\right]\ldots\chi_{i_n}\left[\phi\right] = \\
& = \left.F\left[-i\fundelta{}{J}\right]\exp\left\{i\sum{J_i\chi_i\left[\phi\right]}\right\}\right\vert_{J = 0}\\
&F\left[-i\fundelta{}{J}\right] = \sum(-i)^nF_{i_1\ldots{i_n}}\fundelta{}{J_{i_1}}\ldots\fundelta{}{J_{i_n}}
\end{split}
\end{equation}

A state is a linear functional $\Omega: \mathfrak{U} \rightarrow \mathbb{C}$ such that
\begin{equation}
\Omega\left[\mathbf{1}\right] = 1 \label{formalism:defstate_unity}
\end{equation}
and
\begin{equation}
\Omega\left[A^\dag\right] = \overline{\Omega\left[A\right]} \label{defstate:conjugate}
\end{equation}
The value $\Omega\left[F\right]$ is interpreted as expectation value of the argument.
Therefore, for any \textit{physical} observable $A$ the following nonnegativity condition must hold:
\begin{equation}
\Omega\left[A^{\dag}A\right] \geq 0. \label{formalism:defstate_positivity}
\end{equation}
Not all elements in $\mathfrak{U}$ are required to to be physical though.

Construction of a state starts with fixing the mean field
\begin{equation}
\Omega\left[\phi_a(x)\right] = \varphi_a(x).
\end{equation}
Any $F \in \mathfrak{U}$ is infinitely differentiable by construction. Thus it can be expanded into a functional Taylor series near $\varphi$.
Linearity then implies that 
\begin{equation}
\begin{split}
&\Omega\left[F\right] = \left[ 1 +\right.\\
 &\left. + \sum\limits_{n \geq 2}\idotsint{d^4x_1\ldots{}d^4x_n}\Omega_{a_1\ldots{}a_n}(x_1,\ldots,x_n)\fundelta{}{\phi_{a_1}}\ldots\fundelta{}{\phi_{a_1}}\right]\times\\
 &\left.\times{}F\left[\phi\right]\right\vert_{\phi = \varphi} = 
 \left.F\left[-i\fundelta{}{J}\right]\exp\left(iW\left[\varphi,J\right]\right)\right\vert_{J = 0}
\end{split}\label{formalism:defW}
\end{equation}

\ref{formalism:defW} can be continued to arbitrary values of $J$ by putting
\begin{equation}
\Omega\left[F\right] = \exp\left(-iW\left[\varphi,J\right]\right)F\left[-i\fundelta{}{J}\right]\exp\left(iW\left[\varphi,J\right]\right) \label{formalism:Wext}
\end{equation}
which coincides with \ref{formalism:defW} at $J = 0$. It is obvious that the shift $J \rightarrow J^{\prime}$ can be compensated by adjusting the coefficients in $W\left[J\right]$. 

\ref{formalism:defW} implies that 
\begin{equation}
\begin{split}
&W\left[\varphi,J\right] = W_0\left[\varphi,J\right] + \tilde{W}\left[\varphi, J\right]\\
&W_0\left[\varphi,J\right] = \int{d^4x}J^a(x)\varphi_a(x) - \\ 
&-\frac{1}{2}\iint{}d^4x_1d^4x_2G_{ab}(x_1,x_2)J_\phi^a(x_1)J_\phi^b(x_2) 
\end{split} \label{formalism:splitW}
\end{equation}
where $J_\phi^a(x)$ denotes the source for $\phi_a(x)$. It is easy to see that
\begin{equation}
J_\phi^a(x)e^{iW_0\left[\varphi,J\right]} = \int{d^4y}{G^{-1}}^{ab}(x,y)\left[i\fundelta{}{J_\phi^b(y)} + \varphi_b(y)\right]e^{iW_0\left[\varphi,J\right]}.
\label{formalism:reduction}
\end{equation}
Expanding $\exp\left\{i\tilde{W}\left[\varphi, J\right]\right\}$ and applying \ref{formalism:reduction} repeatedly to eliminate all powers of $J$ yields
\begin{equation}
\exp\left\{iW\left[\varphi,J\right]\right\} = \exp\left\{iV\left[\varphi,\fundelta{}{J}\right]\right\}\exp\left\{iW_0\left[\varphi,J\right]\right\}.
\end{equation}
Since $W_0\left[\varphi,J\right]$ is quadratic, it can be expressed as a Gaussian path integral
\begin{equation}
\begin{split}
&\exp\left\{iW_0\left[\varphi,J\right]\right\} \sim \\
&\sim \funint{\theta}\exp\left\{i\left[\frac{1}{2}\iint{d^4xd^4y}\theta_a(x)G^{-1\,ab}(x,y)\theta_b(y) +\right.\right.\\
&\left.\left.+ \int{d^4x}J^a(x)\left[\varphi_a(x) + \theta_a(x)\right]\right]\right\} 
\end{split}
\end{equation}
where $\sim$ denotes equality up to a $J$-independent multiplier.
Therefore 
\begin{equation}
\begin{split}
&\exp\left\{iW\left[\varphi,J\right]\right\} = \\
&=\funint{\theta}\exp\left\{i\left[\frac{1}{2}\iint{d^4xd^4y}\theta_a(x)G^{-1\,ab}(x,y)\theta_b(y) +\right.\right.\\
&\left.\left.+ \int{d^4x}J^a(x)\left[\varphi_a(x) + \theta_a(x)\right] + V\left[\varphi, \theta\right]\right]\right\} = \\
&=\funint{\phi}\exp\left\{i\Sigma\left[\phi,\varphi,J\right]\right\}\\
&\phi_a(x) = \theta_a(x) + \varphi_a(x)
\end{split} \label{formalism:path_int}
\end{equation}
In other words, any state $\Omega$ can be represented by a path integral. 
Equipping the fields $\phi_a$ with indices $a \in \left\{1,2\right\}$ allows an algebraic definition of closed time path or Schwinger-Keldysh path integral \cite{Schwinger1961}, \cite{Keldysh1964}, \cite{Zhou1980}, \cite{Chou1985}, \cite{Jordan1986}, \cite{CalHu1987}, \cite{Su1988}, \cite{CalHu1988}, \cite{CalHu1989}, \cite{CalHu2008} with a suitable choice of $G$:
$\phi_1$ would belong to the forward-running and $\phi_2$ to the backward-running branch of the integration contour. 
Of course, it is also possible to construct the Feynman path integral along these lines but Schwinger-Keldysh formalism is needed to produce observable expectation values instead of matrix elements \cite[pp. 177--180]{CalHu2008}.

The definition \ref{formalism:Wext} allows to introduce the effective action. Let
\begin{equation}
\begin{split}
&\xi_i = \Omega\left[\chi_i\left[\phi\right]\right] = \\
& = \exp\left(-iW\left[\varphi,J\right]\right)\left[-i\frac{\delta}{\delta{J_i}}\right]\exp\left(iW\left[\varphi,J\right]\right) {}
= \fundelta{W\left[J\right]}{J_i} \label{formalism:mean_xi}
\end{split}
\end{equation} 
Then
\begin{equation}
\Gamma\left[\xi\right] = W\left[J_{\xi}\right] - \sum_{i}J_{\xi_i}(x)\xi_i \label{formalism:effaction}
\end{equation}
here $J_{\xi}$ is such that \ref{formalism:mean_xi} holds for some given $\xi_i$.
Thus
\begin{equation}
\begin{split}
&\fundelta{\Gamma\left[\xi\right]}{\xi_i} = -\sum_k{T_{ik}J_k} \\
&T_{ik} = \fundelta{\xi_k}{\xi_i}\\
&J_i = T^{-1}_{ik}\fundelta{\Gamma}{\xi_k}
\end{split}
\label{formalism:effective_field_eqn}
\end{equation}
This makes it possible to express $\fundelta{J_i}{\xi_k}$ and, by taking the operator inverse, $\fundelta{\xi_i}{J_k}$ through functional derivatives of $\Gamma$. 
As 
\begin{equation}
\fundelta{}{J_k} = \sum_i\fundelta{\xi_i}{J_k}\fundelta{}{\xi_i},
\end{equation}
this allows to express any $\Omega\left[F\right]$ through the effective action.

The place of renormalization in the construction outlined above is best demonstrated by an example. Let 
\begin{equation}
\begin{split}
&W_D\left[J\right] = \iint{d^4x_1d^4x_2}D(x_1 - x_2)J^1(x_1)J^1(x_2)\\
&\left(\square_{x_1} - m^2\right)D(x_1 - x_2) = \delta(x_1 - x_2)
\end{split}
\end{equation}
which corresponds to Klein-Gordon vacuum state.
Then it is easy to see that
\begin{equation}
\begin{split}
&\exp\left(-iW_D\left[J\right]\right)\Phi\left[J\right]\exp\left(iW_D\left[J\right]\right) = \infty\\
&\Phi\left[\fundelta{}{J}\right] = \iiiint{}f(x_1,x_3)\delta(x_1 - x_2)\delta(x_3 - x_4)\times\\
&\times\fundelta{}{J^1(x_1)}\fundelta{}{J^1(x_2)}\fundelta{}{J^1(x_3)}\fundelta{}{J^1(x_4)}{d^4x_1d^4x_2d^4x_3d^4x_4}
\end{split}
\end{equation}
being a sum of two typical loop divergences. 
In general, renormalization is required to allow evaluation of the observables defined on the thin diagonal $x_i = x_j$.
It is done in three steps, according to the BPHZ prescription \cite{Zavyalov1990}, \cite{BoShir1980}:
\begin{enumerate}[label=\arabic*), nosep]
\item define a regularization procedure for loop divergences;
\item insert counterterms into $W$ to subtract divergent parts of basic diagrams;
\item lift the regularization when computation is completed.
\end{enumerate}
If the theory is defined by its classical action, 
BPHZ prescription also requires renormalization to respect its symmetries.
This requirement imposes restrictions on finite renormalization constants arising in the process.

It is worth noting that algebraic approach does not necessarily require to insert counterterms under the path integral.
After applying an intermediate regularization, the following decomposition is possible: 
\begin{equation}
W\left[\varphi,J\right] = W_r\left[\varphi, J\right] + R\left[\varphi,J\right]
\end{equation}
where $W_r\left[\varphi, J\right]$ is the regularized generating functional and $R\left[\varphi,J\right]$ cancels the divergences.
Then the procedure \ref{formalism:splitW} - \ref{formalism:path_int} can be applied to $W_r$ only, leaving $R$ intact.
\ref{formalism:path_int} will then be replaced with
\begin{equation}
\exp\left\{iW\left[\varphi,J\right]\right\} = \exp\left\{iR\left[\varphi,J\right]\right\}\funint{\phi}\exp\left\{i\Sigma\left[\phi,\varphi,J\right]\right\}.
\end{equation}
Thus, renormalization can be applied to the generating functional or to the effective action, bypassing the classical action.

In the simplest and the most widely studied case the propagator $G_{ab}\left(x,y\right)$ in \ref{formalism:splitW} depends on spacetime geometry only 
(in most physical field theories it is possible to separate such purely geometric part and treat the rest perturbatively). 
Therefore, counterterms would also be purely geometric as in the standard BPHZ scheme.
In the general case, however, $\varphi$ cannot be separated and a background-dependent renormalization is needed. This problem would be unavoidable in any attempt to quantize Einstein's gravity.

\section{Effective action for a bosonized NJL-like model} \label{NJL-effaction}

A CP-symmetric model of Nambu-Jona-Lasinio type with additional $U(N)$ global color symmetry will be studied in this section.
Let
\begin{equation}
\begin{split}
&\Gamma_1 = 1\\
&\Gamma^\mu_2 = \gamma^\mu\\
&\Gamma^{\mu\nu}_3 = \sigma^{\mu\nu}\\
&\Gamma^{\mu}_4 = \gamma^5\gamma^\mu\\
&\Gamma_5 = \gamma^5\\
&\tau_{i0}^{\bar{\alpha}} = \Gamma_i^{\bar{\alpha}}\\
&\tau_{i1}^{\bar{\alpha}} = T^a\Gamma_i^{\bar{\beta}},
\end{split} \label{effaction/gammabasis}
\end{equation}
By virtue of Fierz identities for colored fermions \cite{Kutnii2023}, the most general CP-invariant NJL-like Lagrangian is
\begin{equation}
\begin{split}
\Lgr = \bar{\psi}i\widehat{\partial}\psi + 
G_i\bar{\psi}\Gamma_{i\,\bar{\alpha}}\psi\bar{\psi}\Gamma^{\bar{\alpha}}_i\psi
\end{split} \label{effaction/NJL}
\end{equation}
Construction of the effective action starts with the generating functional
\begin{equation}
\begin{split}
Z\left[\left\{J\right\}\right] &= 
\funint{\bar{\psi},\psi}\exp\left\{i\int\left[\bar{\psi}i\widehat{\partial}\psi + 
G_i\bar{\psi}\Gamma_{i\,\bar{\alpha}}\psi\bar{\psi}\Gamma^{\bar{\alpha}}_i\psi + \right.\right.\\
&\left.\left.+ \bar{J}_1\psi +\bar{\psi}J_1 
+ 2\lambda_{\bar{i}}J_{2\bar{i}\bar{\alpha}}\bar{\psi}\tau_{\bar{i}}^{\bar{\alpha}}\psi
+ \lambda_{\bar{i}}J_{2\bar{i}\bar{\alpha}}J^{\bar{\alpha}}_{2\bar{i}}
\right]d^4x\right\},
\end{split} \label{effaction/genfunc}
\end{equation}
which is then bosonized by inserting
\begin{equation}
\begin{split}
&\text{const} = \funint{\Sigma}\exp\left\{-i\int
\left(\frac{\lambda_{\bar{i}}^{-\frac{1}{2}}}{2}\Sigma_{\bar{i}\bar{\alpha}} 
- \lambda_{\bar{i}}^{\frac{1}{2}}\bar{\psi}\tau_{\bar{i}\bar{\alpha}}\psi - \lambda_{\bar{i}}^{\frac{1}{2}}J_{\bar{i}\bar{\alpha}}
\right)\right.\times\\
&\left.\times
\left(\frac{\lambda_{\bar{i}}^{-\frac{1}{2}}}{2}\Sigma_{\bar{i}}^{\bar{\alpha}} 
- \lambda_{\bar{i}}^{\frac{1}{2}}\bar{\psi}\tau_{\bar{i}}^{\bar{\alpha}}\psi - \lambda_{\bar{i}}^{\frac{1}{2}}J_{\bar{i}}^{\bar{\alpha}}
\right)d^4x\right\}. 
\end{split} \label{effaction/bosonization}
\end{equation}
This results in
\begin{equation}
\begin{split}
&Z\left[\left\{J\right\}\right] = 
\funint{\bar{\psi},\psi,\Sigma}\exp\left\{i\int{d^4x}\left[\bar{\psi}\left(i\widehat\partial + \widehat{\Sigma}\right)\psi 
- \frac{\Sigma_{\bar{i}\bar{\mu}}\Sigma_{\bar{i}}^{\bar{\mu}}}{4\lambda_{\bar{i}}} +\right.\right.\\
&\left.\left.+ \bar{J}_1\psi +\bar{\psi}J_1 + J_{2\bar{i}\bar{\mu}}\Sigma_{\bar{i}}^{\bar{\mu}}\right]\right\}\\
&\widehat{\Sigma} =
\Sigma_{10} + T^a\Sigma^a_{11} + \gamma_\mu\Sigma^\mu_{20} + T^a\gamma_\mu\Sigma^{a\mu}_{21} 
+ \sigma_{\mu\nu}\Sigma^{\mu\nu}_{30} + T^a\sigma_{\mu\nu}\Sigma^{a\mu\nu}_{31} +\\
&+ \gamma^5\gamma_\mu\Sigma^\mu_{40} + T^a\gamma^5\gamma_\mu\Sigma^{a\mu}_{41} 
+ \gamma^5\Sigma_{50} + T^a\gamma^5\Sigma^a_{51},
\end{split} \label{effaction/bosonized}
\end{equation}
provided that
\begin{equation}
	\begin{bmatrix}
		G_1\\G_2\\G_3\\G_4\\G_5
	\end{bmatrix} = 
	\begin{bmatrix}
		\lambda_{10}\\\lambda_{20}\\\lambda_{30}\\\lambda_{40}\\\lambda_{50}
	\end{bmatrix} + 
	\begin{bmatrix}
		-\frac{N + 4}{8N}&-\frac{1}{2}&-\frac{3}{2}&\frac{1}{2}&-\frac{1}{8}\\
		-\frac{1}{2}&\frac{N - 2}{4N}&0&\frac{1}{4}&\frac{1}{8}\\
		-\frac{1}{16}&0&\frac{N - 2}{4N}&0&-\frac{1}{16}\\
		\frac{1}{8}&\frac{1}{4}&0&\frac{N - 2}{4N}&-\frac{1}{8}\\
		-\frac{1}{8}&\frac{1}{2}&-\frac{3}{2}&-\frac{1}{2}&-\frac{N + 4}{8N}
	\end{bmatrix}
	\begin{bmatrix}
		\lambda_{11}\\\lambda_{21}\\\lambda_{31}\\\lambda_{41}\\\lambda_{51}		
	\end{bmatrix}.
	\label{effaction/fierz_conditions}
\end{equation}

Now, to compute the effective action using the background field method \cite{Jackiw1974}, the fields must be split into background and deviation parts:
\begin{equation}
\begin{split}
\psi &\rightarrow \Psi + \psi\\
\Sigma_{\bar{i}}^{\bar{\alpha}} &\rightarrow \Sigma_{\bar{i}}^{\bar{\alpha}} + \varsigma_{\bar{i}}^{\bar{\alpha}},
\end{split}
\end{equation}
and the effective action can be written as
\begin{equation}
\begin{split}
&\Gamma\left[\bar{\Psi},\Psi,\Sigma\right] = 
\int\left[\bar{\Psi}\left(i\widehat{\partial} + \widehat{\Sigma}\right)\Psi
- \frac{\Sigma_{\bar{i}\bar{\alpha}}\Sigma_{\bar{i}}^{\bar{\alpha}}}{4\lambda_{2\bar{i}}}
\right]d^4x + W_1\left[\bar{\Psi},\Psi,\Sigma\right] \\
&W_1\left[\bar{\Psi},\Psi,\Sigma\right] = \\
& = - iP_\text{1PI}\ln\funint{\bar\psi,\psi,\varsigma}\exp\left\{i\int\left[
\bar{\psi}\left(i\widehat{\partial} + \widehat{\Sigma}\right)\psi +
\bar{\Psi}\widehat{\varsigma}\psi +\right.\right.\\
&\left.\left.+ \bar{\psi}\widehat{\varsigma}\Psi + \bar{\psi}\widehat{\varsigma}\psi
- \frac{\varsigma_{\bar{i}\bar{\alpha}}\varsigma_{\bar{i}}^{\bar{\alpha}}}{4\lambda_{2\bar{i}}} 
-\fundelta{W_1}{\Psi}\psi - \bar{\psi}\fundelta{W_1}{\bar{\Psi}}
- \fundelta{W_1}{\Sigma_{\bar{i}}^{\bar{\alpha}}}\varsigma_{\bar{i}}^{\bar{\alpha}}
\right]d^4x
\right\}.
\end{split} \label{effaction/defW}
\end{equation}
where $P_\text{1PI}$ means that only 1-particle irreducible diagrams must be taken from the expression to the right of it.

The Lagrangian of deviations in \eqref{effaction/defW} 
\begin{equation}
\Lgr_\text{dev} = \bar{\psi}\left(i\widehat{\partial} + \widehat{\Sigma}\right)\psi +
\bar{\Psi}\widehat{\varsigma}\psi + \bar{\psi}\widehat{\varsigma}\Psi + \bar{\psi}\widehat{\varsigma}\psi
- \frac{\varsigma_{\bar{i}\bar{\alpha}}\varsigma_{\bar{i}}^{\bar{\alpha}}}{4\lambda_{2\bar{i}}}
\end{equation}
is invariant under gauge transformations
\begin{equation}
\begin{split}
&\psi \rightarrow U\psi\\
&\Psi \rightarrow U\Psi\\
&\Sigma \rightarrow U\Sigma{U}^+ - iU^+\widehat{\partial}U\\
&\varsigma \rightarrow U\varsigma{U}^+.
\end{split} \label{effaction/deftransform}
\end{equation}
BPHZ recipe then dictates that gauge symmetry requirements must be imposed on R-operation for $W_1$ order by order. 
As a consequence, R-operation will not affect the vector boson mass term $\frac{\Sigma_{21\mu}\Sigma_{21}^\mu}{4\lambda_{21}}$. 
Thus, vector bosons are born massive in this scheme and there is seemingly no cure for this.

The interaction $\bar{\psi}\widehat{\varsigma}\psi$ can be treated perturbatively, 
and the cross terms $\bar{\Psi}\widehat{\varsigma}\psi + \bar{\psi}\widehat{\varsigma}\Psi$
can be eliminated by a proper shift of integration variables $\bar{\psi}, \psi$
which leads to
\begin{equation}
\begin{split}
&W_1\left[\bar{\Psi},\Psi,\Sigma\right] = 
-i\Tr\ln\left(i\widehat{\partial} + \widehat{\Sigma}\right) + \frac{i}{2}\Tr\ln\left(L + T\right) -\\
&-iP_\text{1PI}\ln\left\{\exp(iV)\times\right.\\&
\left.\times\exp\left[
-iJ_{\Psi}\frac{1}{i\widehat{\partial} + \widehat{\Sigma}}J_{\bar{\Psi}}
+ i\tilde{J}_\Sigma\frac{1}{L + T}\tilde{J}_\Sigma
\right]\right\}\left|\begin{array}{l}
J_\Psi = -\fundelta{W_1}{\Psi}\\
J_{\bar{\Psi}} = -\fundelta{W_1}{\bar{\Psi}}\\
J_\Sigma^{\bar{i}\bar{\alpha}} = -\fundelta{W_1}{\Sigma_{{\bar{i}\bar{\alpha}}}}
\end{array}\right.
\end{split} \label{effaction/loopexpr}
\end{equation}
where
\begin{equation}
\begin{split}
&{L_{\bar{i}\bar{j}}}_{\bar{\alpha}}^{\bar{\beta}}(x,y) = 
\frac{\delta_{\bar{i}\bar{j}}\delta_{\bar{\alpha}}^{\bar{\beta}}}{4\lambda_{2\bar{i}}}\delta(x-y)\\
&{T_{\bar{i}\bar{j}}}_{\bar{\alpha}}^{\bar{\beta}}(x,y) = \bar{\Psi}(x)\tau_{\bar{i}\bar{\alpha}}\mathcal{G}(x,y)\tau_{\bar{j}}^{\bar{\beta}}\Psi(y)\\
&\tilde{J}_{\Sigma\bar{i}\bar{\alpha}} = J_{\Sigma\bar{i}\bar{\alpha}} - J_\Psi\mathcal{G}\tau_{\bar{i}\bar{\alpha}}\Psi 
- \bar{\Psi}\tau_{\bar{i}\bar{\alpha}}\mathcal{G}J_{\bar{\Psi}}\\
&\mathcal{G}(x,y) = \frac{1}{i\widehat{\partial} + \widehat{\Sigma}}\\
&V = i\frac{\delta}{\delta{J_{\bar{\Psi}}}}\tau_{i}^{\bar{\alpha}}
\frac{\delta}{\delta{J_\Psi}}\frac{\delta}{\delta{J_\Sigma^{\bar{i}\bar{\alpha}}}}
\end{split} \label{effaction/notation}
\end{equation}

\section{Generalized NJL model with current metric field} \label{current_metric}

The problem of vector boson masslessness can be solved by promoting interaction constants in \eqref{effaction/NJL}
to a field. A realistic model must also include parity-breaking interactions. Let
\begin{equation}
T^{\bar{a}} \in \left\{T^a_L \equiv \frac{1 - \gamma^5}{2}T^a, T^a_R \equiv \frac{1 + \gamma^5}{2}T^a\right\}. \label{current_metric/defalgebra}
\end{equation}

A generating functional of the form
\begin{equation}
\begin{split}
&Z\left[\{J\}\right] = \funint{\bar{\psi},\psi,\Lambda}\exp\left\{i\int\left[
\bar{\psi}i\widehat{\partial}\psi +\right.\right.\\
&+ \left(\frac{\zeta^2}{\Lambda^2}\right)^{\bar{a}\bar{b}}\bar{\psi}\gamma_{\mu}T^{\bar{a}}\psi\bar{\psi}\gamma^{\mu}T^{\bar{b}}\psi 
+ G_i\bar{\psi}\Gamma_{i\bar{\alpha}}\psi\bar{\psi}\Gamma_i^{\bar{\alpha}}\psi +\\
&\left.\left.+ H_i\bar{\psi}\Gamma_{i\bar{\alpha}}\psi\bar{\psi}\gamma^5\Gamma_i^{\bar{\alpha}}\psi + \Lgr(\Lambda) + \Lgr_\text{src}(\{J\}, \bar{\psi},\psi,\Lambda)\right]d^4x\right\} 
\end{split} \label{current_metric/genfunc}
\end{equation}
will be studied, where $\Lambda^{\bar{a}\bar{b}}$ defines a symmetric linear form on the algebra \eqref{current_metric/defalgebra},
and $\Lgr_\text{src}(\{J\}, \bar{\psi},\psi,\Lambda)$ is the sum of source terms.
As $\Lambda^{-2}$ defines coupling between fermion currents, it can be called current metric. 

The $\Lambda-\psi$ interaction can be eliminated by inserting
\begin{equation}
\begin{split}
&\text{const} = \funint{A}\exp\left\{\Tr\ln{M_A} -\right.\\
&\left.-i\int\left[\frac{1}{2}A_\mu^{\bar{a}}M_A^{\bar{a}\bar{b}} - \bar{\psi}T^{\bar{a}}\gamma_\mu\psi{}M_A^{-1\,\bar{a}\bar{b}} - J_{A\mu}^{\bar{a}}M_A^{-1\,\bar{a}\bar{b}}\right]\times\right.\\
&\left.\times\left[\frac{1}{2}M_A^{\bar{b}\bar{c}}A_\mu^{\bar{c}} - M_A^{-1\,\bar{b}\bar{c}}\bar{\psi}T^{\bar{c}}\gamma_\mu\psi - M_A^{-1\,\bar{b}\bar{c}}J_{A\mu}^{\bar{c}}\right]d^4x\right\},
\end{split}
\end{equation}
where $M_A$ is defined by
\begin{equation}
\begin{split}
&\left[M_A^{-2}\right]^{\bar{a}\bar{b}}\bar{\psi}\gamma_{\mu}T^{\bar{a}}\psi\bar{\psi}\gamma^{\mu}T^{\bar{b}}\psi =\\
&=
\left(\frac{\zeta^2}{\Lambda^2}\right)^{\bar{a}\bar{b}}\bar{\psi}\gamma_{\mu}T^{\bar{a}}\psi\bar{\psi}\gamma^{\mu}T^{\bar{b}}\psi 
+ K_{LL}\bar{\psi}\gamma_{\mu}T^a_L\psi_L\bar{\psi}\gamma^{\mu}T^a_L\psi +\\
&+ K_{LR}\bar{\psi}\gamma_{\mu}T^a_L\psi\bar{\psi}\gamma^{\mu}T^a_R\psi
+ K_{RR}\bar{\psi}\gamma_{\mu}T^a_R\psi\bar{\psi}\gamma^{\mu}T^a_R\psi
\end{split} \label{current_metric/mass_func}
\end{equation}
Here $K_{\chi_1\chi_2}$ are arbitrary real constants. By virtue of Fierz identities, adding the $K$-terms alters $G_i,H_i$.

Bosonization of remaining fermion interactions looks essentially the same as in the previous section.
With a suitable choice of $\Lgr_\text{src}$, the bosonized generating functional will have the form
\begin{equation}
\begin{split}
&Z\left[\{J\}\right] = \funint{\bar{\psi},\psi,\Lambda,A}\exp\left\{i\int\left[
\bar{\psi}\left(i\widehat{\partial} + \widehat{A} + \widehat{\Sigma}\right)\psi +\right.\right.\\
&+ \bar{J}_\psi\psi + \bar{\psi}J_\psi 
- \frac{M_A^{2\,\bar{a}\bar{b}}(\Lambda)}{4}A_\mu^{\bar{a}}A^{\bar{b}\mu} 
- \frac{1}{2}M^2_{\Sigma\,\bar{i}}\Sigma_{\bar{i}\bar{\alpha}}\Sigma_{\bar{i}}^{\bar{\alpha}} +\\
&\left.\left.+ J_{A\mu}^{\bar{a}}A^{\bar{a}\mu} +J_{\Sigma\bar{i}\bar{\alpha}}\Sigma_{\bar{i}}^{\bar{\alpha}}
+ \Lgr(\Lambda) + J_{\Lambda}^{\bar{a}\bar{b}}\Lambda^{\bar{b}\bar{a}}\right]d^4x + \Tr\ln{M_A}\right\} 
\end{split} \label{current_metric/V-func}
\end{equation}

As it was discussed in the previous section, the fermion loop corections will not affect the $A$-field's mass matrix but $\Lambda$ loop terms will.
Thus, if $\Lambda$ has a nontrivial ground state average $\Lambda_g$, the mass matrix will have the form
\begin{equation}
\mathcal{M}_A = \left[\zeta^2 + \Lambda_g^2K\right]^{-1}\Lambda_g^2 + \mathcal{M}_{A\text{loop}}(\Lambda_g, K, \zeta), 
\end{equation}
where $K$ is the matrix of coefficients $K_{\chi_1\chi_2}$ introduced in \eqref{current_metric/mass_func}.
As $K$ is arbitrary, it seems plausible for a wide class of Lagrangians $\Lgr(\Lambda)$ that
$\mathcal{M}_A$ can be rendered degenerate but not zero by adjusting $K$.
In this case, some components of $A$ will become massless while others might gain large non-Higgs masses.
This makes the bosonized model a candidate grand unified theory. 
As vector boson masses can be of partially non-Higgs origin in this model, the requirements for Higgs sector may be less stringent than in gauge GUTs.
An $U(24)$ model with fermions bearing a single color-isospin-quark/lepton-generation index should be able to produce observable physics.
It is worth noting that an attempt to implement generation as a second index labelling fermion types but taking no part in the interactions would not succeed 
as Fierz transformations can produce generation-changing interactions anyway.

\section{Discussion}

In certain aspect, mass generation in bosonized NJL-like models is the opposite of Higgs mechanism: 
while the Higgs field gives masses to originally massless gauge bosons, 
the current metric field is needed to make vector bosons \textit{massless}. 
The algebraic formalism developed in the section \ref{formalism} 
helps to clarify the role of Fierz coefficients $K_{\chi_1\chi_2}$ 
which is a total mystery in the path integral approach.

In the algebraic picture, a state $\Omega$ is essentially a collection of fundamental correlators 
from which expectation values of observables can be constructed. 
It is clear that those correlators depend on the choice of $K_{\chi_1\chi_2}$. 
Thus, $K_{\chi_1\chi_2}$ implicitly define the family of states represented by the path integral \ref{current_metric/V-func}.
The physical question that remains open here is why such a peculiar family of states with massless vector bosons is preferred by nature.

As the model \ref{current_metric/genfunc} produces higgsoid fields naturally alongside the vector bosons,
its predictions can be not so different from those of gauge GUTs. 
In principle, the case with $\mathcal{M}_A = 0$ and all vector boson masses coming from the Higgs sector is possible.
However, the degenerate $\mathcal{M}_A$ case with both Higgs and non-Higgs contributions to particle masses is more interesting.
While photons produced by such a model must be massless, and gluons should most likely ne massless,
there is no obvious reason why weak bosons' masses cannot have a non-Higgs part.

The lack of a local gauge invariance in the current metric field model has one interesting implication: 
't Hooft-Polyakov monopoles become unlikely. 
As field configuration in a monopole approaches pure gauge asymptotically at spatial infinity \cite{Polyakov1974}, 
presence of noninvariant terms in the effective action would make its energy infinite unless some magical cancellation happens.
This could explain why magnetic monopoles are not observed.

However, the possibility to produce a reasonable GUT is not the only interesting aspect of NJL-like models studied here.
Being nonrenormalizable and, in the case of current metric field theory, non-polynomial, they pose a number of formal challenges
responding to which urges to rethink the mathematical methods of quantum field theory.

The traditional approach is to start with a classical model and quantize it. 
While natural in quantum electrodynamics which is meant to produce small corrections to the classical picture, 
it is much less so in the Standard Model and especially when unified models are considered: 
most fields present in the Lagrangian are not observable directly. 
Thus the classical models hardly correspond to any physical reality.
As the distance between the phenomenology and the classical Lagrangian grows, 
the latter becomes a feature of the formalism. 
It then becomes tempting to bypass the classical level altogether and learn to construct effective actions directly from first principles.
Should this program succeed, it would be comparable to becoming native quantum language speakers 
instead of translating from classical to quantum all the time.
The path of algebraic quantum field theory seems to lead in this direction.

\section{Summary and conclusions}
Bosonization in NJL-like models can be understood as a version of effective action formalism with composite fields.
Formal challenges arising from nonrenormalizability of the model can be addressed by redefining the formalism in the language of algebraic quantum field theory. This allows to develop BPHZ renormalization for the effective action. Symmetries of the effective action imply that vector bosons are naturally massive in a plain NJL-like model under BPHZ renormalization. Modifying the model by introduction of the current metric field can produce massless vector bosons. This is likely true for a wide class of current metric Lagrangians and does not depend on the approach to divergences.
Since bosonization produces Higgs-like fields as well, NJL-like models with current metric field 
open a number of possibilities to produce realistic particle phenomenology. 
There can be both Higgs and non-Higgs contributions to particle masses in such models.
Non-gauge nature of these theories offers a natural explaination for absence of magnetic monopoles in the experiment.

\bibliographystyle{elsarticle-harv}
\bibliography{current_metric} 

\end{document}